\documentclass[12pt]{article}
\usepackage{graphicx}
\usepackage{psfig,epsfig}
\def\Frac#1#2{\frac{\displaystyle{#1}}{\displaystyle{#2}}}
\def\lsim{\raise0.3ex\hbox{$\;<$\kern-0.75em\raise-1.1ex\hbox{$\sim\;$}}}
\def\gsim{\raise0.3ex\hbox{$\;>$\kern-0.75em\raise-1.1ex\hbox{$\sim\;$}}}

\def\app#1#2#3{ Astroparticle Phys. {\bf #1} (#2) #3}

\def\apj#1#2#3{ Astrophys. J. {\bf #1} (#2) #3}

\def\np#1#2#3{ Nucl. Phys. {\bf #1} (#2) #3}
\def\nat#1#2#3{ Nature {\bf #1} (#2) #3}
\def\pl#1#2#3{ Phys. Lett. {\bf #1} (#2) #3}
\def\pr#1#2#3{ Phys. Rev. {\bf #1} (#2) #3}
\def\prep#1#2#3{ Phys. Rep. {\bf #1} (#2) #3}
\def\prl#1#2#3{ Phys. Rev. Lett. {\bf #1} (#2) #3}

\def\jhep#1#2#3{ J. High Energy Phys. {\bf #1} (#2) #3}
\def\jmps#1#2#3{ J. Moscow Phys. Soc. {\bf #1} (#2) #3}

\def\rpp#1#2#3{ Rept. Prog. Phys. {\bf #1} (#2) #3}

\newcommand{\be}{\begin{equation}}
\newcommand{\ee}{\end{equation}}

\newcommand{\bea}{\begin{eqnarray}}
\newcommand{\eea}{\end{eqnarray}}
\newcommand{\eqn}[1]{(\ref{#1})}
\newcommand{\pp}{~~~.}
\newcommand{\vv}{~~~,}

\newcommand{\gapproxeq}
{\lower .7ex\hbox{$\;\stackrel{\textstyle >}{\sim}\;$}}
\newcommand{\lapproxeq}
{\lower .7ex\hbox{$\;\stackrel{\textstyle <}{\sim}\;$}}
\begin{document}
%\thispagestyle{empty}
%\begin{titlepage}
\begin{center}
\hfill SISSA 51/2000/EP\\
\hfill DSF 16/2000\\
\vskip 0.5cm
{\Large \bf Non equilibrium spectra of degenerate relic neutrinos}
\end{center}
\normalsize
\vskip1cm
\begin{center}
\baselineskip=13pt
{\bf S. Esposito$^{~a}$, G. Miele$^{~a}$, S. Pastor$^{~b}$
\footnote{Corresponding author. Tel: +39-0403787478. Fax:
+39-0403787528. E-mail: pastor@sissa.it}, M. Peloso$^{~b}$,
O. Pisanti$^{~a}$}\\
\end{center}
\begin{center}
\baselineskip=13pt
{\sl $^a$ Dipartimento di Scienze Fisiche}\\
\baselineskip=12pt
{\sl Universit\`{a} di Napoli {\it Federico II}
\& I.N.F.N., Sezione di Napoli
\\ Complesso Universitario di Monte Sant'Angelo\\ Via Cintia, 80126 Napoli,
Italy}\\
\vglue 0.6cm
\baselineskip=13pt
{\sl $^b$ S.I.S.S.A. and I.N.F.N., Sezione di Trieste}\\
\baselineskip=12pt
{\sl Via Beirut 2-4, 34014 Trieste, Italy}\\
\vglue 0.8cm
\end{center}

\vskip.5cm
\begin{abstract}
We calculate the exact kinetic evolution of cosmic neutrinos until
complete decoupling, in the case when a large neutrino asymmetry
exists. While not excluded by present observations, this large
asymmetry can have relevant cosmological consequences and in
particular may be helpful in reconciling Primordial Nucleosynthesis
with a high baryon density as suggested by the most recent
observations of the Cosmic Microwave Background Radiation. By solving
numerically the Boltzmann kinetic equations for the neutrino
distribution functions, we find the momentum-dependent corrections to
the equilibrium spectra and briefly discuss their phenomenological
implications.
\end{abstract}


\begin{section}{Introduction}

It is a common point of view to treat as instantaneous several
(prolonged) processes that occurred in the Universe during the first
instants after the Big Bang. One the most striking examples is
provided by reheating after inflation. In this case, the approximation
of treating all inflaton quanta as decaying at the same time
$1/\Gamma$ (with $\Gamma$ denoting the inflaton decay rate) is useful
for many purposes, such as the computation of the reheating
temperature $T_R\,$, but it may fail to reproduce the abundances of
the relic particles, especially if they have masses larger than $T_R$
and have never been in equilibrium with the thermal bath \cite{TONI}.

Even when we consider the more common situation of a particle which was
first in equilibrium with the radiation and then decouples, a detailed
study of the whole process may provide more accurate results with
respect to the standard computation, which assumes instantaneous
decoupling. A particularly interesting case is that of neutrinos, which
were coupled to the thermal bath via the well known weak interactions.

Let us first consider the standard situation, where neutrinos are
assumed to decouple when the temperature of the primordial universe is
about $2 - 3$ MeV \cite{KOLTUR}, namely when the rates of the weak
interactions which couple them to the electromagnetic plasma become
smaller than the Hubble parameter. After this instantaneous
decoupling, the neutrino spectra maintain their equilibrium shape, with
the temperature redshifted as the inverse scale factor, $T_\nu \sim
1/R \left( t \right)\,$, due to the expansion of the Universe. In the
meantime, also the temperature of the electromagnetic bath scales in
the same way, until it reaches the electron mass $T_\gamma = m_e
\simeq 0.5$ MeV. At this stage the $e^+ \, e^- \rightarrow \gamma \,
\gamma$ annihilations occur, without affecting the relic neutrinos
previously decoupled. As a consequence, the temperature of photons
increases with respect to $T_\nu$ until it reaches the well--known
asymptotic ratio $T_\gamma / T_\nu = \left( 11/4 \right)^{1/3} \simeq
1.401$.

Relaxing the assumption of instantaneous neutrino decoupling, the
above results slightly modify. The main physical reason is that the
neutrino plasma receives a small contribution from the $e^+ \, e^-$
annihilations, and its final energy density is a little bit higher
than in the standard case. Neutrinos with higher momenta are more
heated since, in the range of energies we are interested in, weak
interactions get stronger with rising energy. This causes a
momentum-dependent distortion in the neutrino spectra from the pure
equilibrium Fermi--Dirac shape.

There have been a number of papers which considered the effects of
non-instantaneous neutrino decoupling
\cite{DICUS,HERRERA,RANA,DT,DF,MADSEN,DHS}. The first papers used
integrated quantities to estimate that the neutrino energy density
increases by a factor $1\%$ \cite{DICUS,HERRERA,RANA}, while
subsequent works \cite{DT,DF} made some momentum-dependent
calculations assuming Boltzmann statistics for neutrinos and other
approximations.  The full numerical computation of the evolution of
the neutrino distribution functions without approximations requires
the numerical solution of the Boltzmann equations, as done in
\cite{MADSEN,DHS}. The most accurate results for the evolution of the
neutrino distortion until complete decoupling are given in
Ref.~\cite{DHS}.  For what concerns the energy stored in the relic
neutrinos, this study gives a temperature ratio after decoupling
$T_\gamma / T_\nu = 1.3991 \,$, which indeed shows that neutrinos also
share a small part of the energy transfer from the $e^+ \, e^-$
annihilations. The method adopted in this computation has however been
questioned in Ref.~\cite{Gnedin} (see also \cite{replyDHS} for a reply
to this work). One of the aims of our work is to verify the results of
Ref.~\cite{DHS}. We have also performed a full numerical calculation,
although employing a different method. Our analysis confirms the
results of Ref.~\cite{DHS}.

The main goal of our study is to extend the previous analyses to the
case in which a large asymmetry between neutrinos and antineutrinos
is present.  If this asymmetry is of order one, one says that
neutrinos are {\it Fermi--degenerate}. Big Bang Nucleosynthesis (BBN)
forces the baryonic asymmetry $\eta_b \equiv \left( n_b - n_{\bar b}
\right) / n_\gamma$ to be very tiny (of the order $10^{-\,10} -
10^{-\,9}$) \cite{bbn}, and the observed electric neutrality of the
Universe translates these bounds on electrons too. However, the
existence of a large asymmetry (even of order one) in the neutrino
sector is an open possibility which has drawn much attention in the
past.

Actually, there exist some realistic theoretical models that can
generate very large values of the lepton asymmetry
$\eta_{\nu}$. {}From a particle physics point of view, a lepton
asymmetry can be generated by an Affleck--Dine mechanism \cite{AF}
without producing a large baryon asymmetry (see
Refs.~\cite{DK,DolgovRep,Casas,MMR,McDonald}), or even by
active--sterile neutrino oscillations \cite{Foot,Pastor} after the
electro--weak phase transition. In general, the asymmetry is expected
to be different for each neutrino family.

Important bounds on $\eta_{\nu}$ come again from BBN (see
\cite{revSarkar} for a review of non-standard BBN). The asymmetry on
electronic neutrinos is the most constrained, since it has a direct
influence on the weak interactions between protons and neutrons and
thus affects the final $^4$He abundance. Asymmetry on the neutrinos of
the two other families is instead less bounded, since its only effect
is to increase the expansion rate of the universe. The simultaneous
presence of an asymmetry in the three neutrino families can still lead
to a successful prediction of the abundances of light elements (fully
degenerate BBN), provided that their values are suitably chosen.  This
last possibility usually requires the presence of a baryon asymmetry
somewhat larger than the one allowed by standard nucleosynthesis (that
is with non-degenerate neutrinos). Indeed, a high neutrino asymmetry
can be considered the simplest option to save BBN predictions in case
a high baryon asymmetry is required. The very recent Boomerang
\cite{boom} and Maxima \cite{maxima} results on the acoustic peaks of
the Cosmic Microwave Background Radiation (CMBR) seem to favour
\cite{boomana1,boomana2,boomana3,Jul} high values for $\eta_b$, which
may be in conflict with the bounds coming from standard BBN. In
particular, in Ref.~\cite{Jul} it is argued that the presence of a
neutrino asymmetry may contribute to both (i) improve the fits of the
Boomerang data and (ii) render the high $\eta_b$ needed compatible
with (degenerate) BBN.

Finally, a relic neutrino asymmetry, delaying the matter domination
stage, can have very significant effects on the matter power
spectrum. In particular, it can suppress the power at small with
respect to large scales, thus making the predictions of Cold Dark
Matter models compatible with observations.

Motivated by these considerations, in this work we extend the study of
non-equilibrium effects on the relic neutrino spectra to the case of
non-vanishing neutrino chemical potentials (degenerate neutrinos). The
paper is organized as follows. In section 2 we review the effect of
neutrino chemical potentials on BBN and other cosmological
observables. The set of equations ruling the evolution of neutrino
distributions and the method adopted to get the solution are presented
in section 3. Numerical details are given in section 4, whereas
section 5 contains our results.  We discuss in section 6 some
phenomenological implications of the non-equilibrium effects. Finally,
in section 7 we give our conclusions.
\end{section}

\begin{section}{Cosmological implications of degenerate neutrinos}

The existence of a relic neutrino asymmetry would have important
cosmological consequences mainly on Primordial Nucleosynthesis and the
CMBR and matter power spectra.

Degenerate neutrinos influence BBN in two distinct ways. The first one
is connected to the increase of the energy density of the primordial
plasma due to a non-vanishing neutrino asymmetry. One can introduce an
{\it effective} number of relativistic neutrinos $N_\nu$ which, for
degenerate neutrinos, depends on the chemical potential of any
neutrino flavour.  The quantity $N_\nu$ is defined, in fact, from the
total neutrino (plus antineutrino) energy density, $\rho_\nu$, through
the relation
\be
\rho_\nu = N_\nu \, \frac{7}{4} \, \frac{\pi^2}{30} \, T_\nu^4 \vv
\label{neff}
\ee
which, for three massless neutrinos with chemical potentials
$\mu_{\nu_\alpha}$ and in the equilibrium case, becomes
\be
N_{\nu} = 3 + \sum_{\alpha =e,\mu, \tau} \left[ \frac{15}{7} \left(
\frac{\xi_\alpha}{\pi} \right)^4 + \frac{30}{7} \left(
\frac{\xi_\alpha}{\pi} \right)^2 \right] \vv
\label{neff1}
\ee 
where $\xi_\alpha = \mu_{\nu_\alpha}/T_\nu$ are the degeneracy
parameters.  Note that $N_{\nu}$ does not depend on the sign of
$\xi_\alpha$. For what concerns BBN, $N_{\nu}>3$ leads to a higher
neutron to proton ratio since it favours an earlier decoupling of
nucleons. This produces an increase in the final production of both
$^4$He (because there are more neutrons available when the nuclei
form), D and $^3$He (since there is less time to destroy them).

A second important effect on BBN predictions is induced by the
asymmetry for $\nu_e$ only, since they directly participate in the $n
\leftrightarrow p$ weak processes. An excess of electronic neutrinos
over antineutrinos ($\xi_e > 0$), in fact, enhances the $n \rightarrow
p$ conversion rate with respect to the inverse process, so reducing
the neutron to proton ratio. In addition, the initial $n/p$ ratio
diminishes by a factor $\exp(-\xi_e)$, and this further reduces the
number of neutrons available at the onset of BBN.  As a consequence,
the observed abundances of light elements can be achieved also in this
case, but with a value of $\Omega_b h^2$ significantly higher than for
$\xi_e=0$.

When more than one neutrino species is degenerate, both the above
effects combine and, as a result, one can observe particular
combinations of values of $\xi_\alpha$ for which the predictions of
degenerate BBN are still in good agreement with the observational data
on the abundances of primordial elements.

An exhaustive analysis of degenerate BBN was performed in
Ref.~\cite{Kang} (see refs.~therein for previous works), while more
recently some aspects have been studied in \cite{KKS,WS,japan}.
Neglecting the non-electron neutrino chemical potentials, the authors
of \cite{Kang} find the limits $-0.06 \leq \xi_e \leq 0.14$.  Instead,
for fully degenerate BBN and requiring that the Universe had a
sufficiently extended period of matter domination, the neutrino
degeneracy parameters lie in the wider ranges \cite{Kang}
\be
-0.06 \leq \xi_e \leq 1.1 ~~~ \mbox{and} ~~~
\left\{ 
\begin{array}{cl}
|\xi_{\mu, \tau}| \leq 6.9 & 
\mbox{for}~~\xi_\mu \neq 0~, ~\xi_\tau=0~\mbox{(or vice versa)}\\
|\xi_{\mu, \tau}| \leq 5.6 & \mbox{for}~~\xi_\mu =\xi_\tau \neq 0
\end{array}
\right.
\label{KSlimits}
\ee
since, as mentioned above, at least for $^4$He, the effect of a
positive $\xi_e$ can be compensated by the contribution to $N_\nu$
coming from $\xi_{\mu,\tau}$.

In a very recent work \cite{napolibbn}, the BBN bounds on the neutrino
degeneracies have been re-analyzed. The input parameters of
degenerate BBN are $\xi_e$, $N_\nu (\xi_\alpha)$ and $\eta_{10} \equiv
10^{10}\eta_b$. Using an updated code, which includes all relevant
physical effects that influence the $^4$He abundance up to $0.1\%$,
one can perform a likelihood analysis of compa\-ti\-bility between
theoretical predictions and experimental data. This analysis yields
contour levels which are surfaces in the three-dimensional space of
these parameters.  However, one should take into account other non BBN
constraints for reducing the ranges of the parameters. For example,
considering the upper bound on the radiation density present at
recombination coming from CMBR data, ref.~\cite{Hannestad} obtains a
$2\sigma$ limit $N_\nu < 13$.  As far as $\xi_e$ and $\eta_{10}$ are
concerned, the analysis is limited to the range $-1 \div 1$ and $1
\div 30$, respectively. The maximum for these functions is found for
\cite{napolibbn}
\be
\label{maxdl}
\xi_e = 0.06 \vv \qquad N_\nu = 3.43 \vv \qquad \eta_{10} = 5 \vv
\ee
for a low value of the D abundance, and
\be
\label{maxdh}
\xi_e = 0.35 \vv \qquad N_\nu = 13 \vv \qquad \eta_{10} = 4.20 \vv
\ee
in the high D case (see figures 19 and 20 of \cite{napolibbn} for the
$95 \%$ exclusion plots for the $\xi_e$ and $N_\nu$ parameters for
different values of $\eta_{10}$). For low D the allowed range for
$\eta_{10}$ is $3.3 \div 9.9$, while for high D we have $1.1 \div
5.8$. 

Primordial nucleosynthesis is not the only framework which can be
substantially affected by a non--zero relic neutrino degeneracy. For
example, in the recent past, several papers have considered the
imprint of this asymmetry on the power spectra of CMBR anisotropies
and matter density. It has been found \cite{Sarkar,Kinney,Lesgourgues}
that the asymmetry boosts the amplitude of the first CMBR peak, shifts
the peaks to larger multipoles, and suppresses small scale matter
fluctuations (see \cite{Larsen} for a previous work). All these
effects are the consequences of increasing the neutrino energy
density, which delays the epoch of matter-radiation equality, and from
their analysis one can put constraints on the neutrino degeneracy
\cite{Sarkar,Lesgourgues}. Among the most recent observations, some
interesting features have emerged from the new Boomerang results
\cite{boom}, which show a quite puzzling suppression of the second
acoustic peak with respect to the first one. A possible explanation to
this problem is provided by the simultaneous presence of both high
baryonic and high neutrino asymmetries, as shown in
Ref. \cite{Jul}. The possible consequences of the neutrino degeneracy
on the future CMBR measurements (Planck) are discussed in
Ref. \cite{Kinney}.

As far as the matter power spectrum is concerned, it is shown in Ref.
\cite{Lesgourgues} that a non-vanishing neutrino degeneracy can
suppress the small scale fluctuations, so as to agree with the
observations even in a pure Cold Dark Matter scenario. It is also
shown in Ref. \cite{Lesgourgues} that the suppression at small scales
is particularly efficient if the degenerate neutrinos are massive,
because free-streaming of non-relativistic neutrinos is enhanced when
their average momentum is boosted by the chemical potential.

The existence of a relic lepton asymmetry enhances the contribution of
massive neutrinos to the present energy content of the Universe
\cite{Pal,geku,Sarkar,Lesgourgues}. Actually it has been shown that
even the smallest neutrino mass suggested by Super--Kamiokande data on
atmospheric neutrinos \cite{SK} could be extracted from CMBR
anisotropy and large-scale structure data by the future Planck
satellite and Sloan Digital Sky Survey, provided that a large neutrino
asymmetry exists \cite{Prunet}.

Finally, it is worthwhile observing that there could be other
implications of the presence of degenerate relic neutrinos, that
include the explanation \cite{geku} of the ultra-high energy cosmic
rays, beyond the Greisen-Zatsepin-Kuzmin cut-off of about $5\times
10^{19}$ eV.  These cosmic rays would be produced by the protons from
the annihilation of ultra-high energy neutrinos on the relic
degenerate neutrinos in the galactic halo (more abundant than in the
standard case) at energies close to the $Z$-resonance.
\end{section}

\begin{section}{The dynamics of neutrino distributions}

At the time of neutrino decoupling, the evolution of their
distributions functions is described by a set of Boltzmann equations
where the collisional terms are due to the weak interactions of
neutrinos with the primordial plasma. Since the baryonic component is
much smaller than the leptonic one, we can safely neglect its
presence. In this case, the whole set of relevant reactions are those
reported in Ref.~\cite{DHS}, where a complete study of the evolution
of neutrino distributions for the non-degenerate case, namely
vanishing chemical potentials, is performed. In this section we
describe a method which allows us to extend the results of
Ref.~\cite{DHS} to the degenerate case.

Following the notations of Ref.~\cite{DHS}, we choose as {\it time}
variable $x \equiv m_e \, R$, and comoving momentum $y \equiv p_\nu\, R$,
where $R$ is the universe scale factor and $m_e$ is the electron mass. We
also define the {\it rescaled} photon temperature as $z \equiv T_\gamma R$.

At sufficiently high temperatures neutrinos can be considered in
thermal equilibrium with the $\gamma$, $e^{\pm}$ plasma through weak
interactions (we always consider temperatures below the muon
mass\footnote{In the range of $\xi_\nu$ we are interested in the
$\nu\bar{\nu}$ annihilation gets out of equilibrium at temperatures
well below the muon mass \cite{Kang,Freese}.}).  Thus they are described
by Fermi-Dirac distributions with the same temperature of the
electromagnetic plasma. However, at lower temperature one expects a
different evolution for the distributions of electronic neutrinos and
antineutrinos, which also experience charged current interactions due
to the presence of $e^{\pm}$ in the thermal bath, with respect to the
neutrinos of the other families. Thus, in the following, we will
assume identical distributions for muon and tau neutrinos
(antineutrinos),
\be f_{\nu_\mu}=f_{\nu_\tau} \equiv f_{\nu_x}(x,y) \vv
~~~ f_{{\bar \nu}_\mu}=f_{{\bar \nu}_\tau} \equiv f_{{\bar
\nu}_x}(x,y) \vv 
\ee 
so restricting the unknown neutrino distributions to $f_{\nu_e}$,
$f_{\bar{\nu}_e}$, $f_{\nu_x}$ and $f_{\bar{\nu}_x}$ only.

In the temperature range we are interested in, electrons and positrons are
kept in thermodynamical equilibrium with photons by fast electromagnetic
interactions. Thus, they are distributed according to the Fermi function
\footnote{We neglect the completely irrelevant $e^{\pm}$ asymmetry since it
is expected to be $\xi_e \sim \eta_b \leq 10^{-\,9}\:$.}
\be
f_{e^-}(x,y,z)=f_{e^+}(x,y,z) = \frac{1}{\exp(\sqrt{y^2+x^2}/z)+1} \pp
\ee

In order to get the {\it time} evolution of neutrino distributions,
$f_{\nu_\alpha}(x,y)$, and the rescaled photon temperature, $z(x)$, one
must solve the following set of differential equations
\bea
\frac{d}{dx} \bar{\rho}(x) &=& \frac 1x\, \left(\bar{\rho}-3\bar{P}
\right)_{m} \vv \label{energy2} \\
\frac{d}{d x}\, f_{\nu_\alpha}(x,y) &=& \frac{1}{x H}\,
I_{\nu_\alpha} \left[ f_{\nu_e},f_{\bar{\nu}_e},f_{\nu_x},f_{\bar{\nu}_x}
\right] \vv ~~~~ \mbox{with} ~~~ \nu_\alpha=\nu_e, \bar{\nu}_e, \nu_x,
\bar{\nu}_x \pp
\label{boltz}
\eea
In equation \eqn{energy2}, which states the conservation of the total
energy density, $\bar{\rho}$ and $\bar{P}$ are the dimensionless
energy density and pressure of the primordial plasma, respectively,
\be \bar{\rho} = \rho \left (\frac{x}{m_e} \right)^4 \vv \qquad
\bar{P} = P \left (\frac{x}{m_e} \right)^4 \vv 
\ee
and the index $m$ in the r.h.s.~reminds that only massive components
in the plasma contribute. {}From \eqn{energy2}, by using the expressions
\bea
\bar{\rho}_\gamma &=& \frac{\pi^2}{15} z^4 \vv \label{rhog} \\
\bar{\rho}_e &=& \frac{2}{\pi^2} \int_0^\infty dy ~y^2
~\frac{\sqrt{x^2+y^2}}{\exp(\sqrt{y^2+x^2}/z)+1} \vv \label{rhoe} \\
\bar{P}_e &=& \frac{2}{3\, \pi^2} \int_0^\infty dy~
\frac{y^4}{\sqrt{x^2+y^2}} ~\frac{1}{\exp(\sqrt{y^2+x^2}/z)+1} \vv
\label{pe} \\
\bar{\rho}_\nu &=& \frac{1}{2\pi^2} \int_0^\infty dy ~y^3 ~\left[
f_{\nu_e}(x,y) + f_{\bar{\nu}_e}(x,y) + 2\, f_{\nu_x}(x,y) + 2\,
f_{\bar{\nu}_x}(x,y) \right] \vv \label{rhonu}
\eea
we get the equation for the evolution of $z(x)$,
\be
\frac{dz}{dx} = \Frac{\frac{x}{z} F_1 (x/z) - \frac{1}{4z^3} \int_0^\infty~
dy~ y^3~ \left (\frac{d f_{\nu_e}}{dx}+ \frac{d f_{\bar{\nu}_e}}{dx}+ 2\,
\frac{d f_{\nu_x}}{dx}+ 2\, \frac{d f_{\bar{\nu}_x}}{dx} \right )}
{\frac{x^2}{z^2} F_1 (x/z) + F_2 (x/z) + \frac{2\pi^4}{15}} \vv
\label{dzdx}
\ee
where the functions $F_i$ are given by
\bea
F_1 (\tau) & \equiv & \int_0^\infty~ d\omega~ \omega^2~
\frac{\exp(\sqrt{\omega^2+\tau^2})}
{(\exp(\sqrt{\omega^2+\tau^2})+1)^2} \vv \label{functionF1} \\
F_2 (\tau) & \equiv & \int_0^\infty~ d\omega~ \omega^4~
\frac{\exp(\sqrt{\omega^2+\tau^2})} {(\exp(\sqrt{\omega^2+\tau^2})+1)^2}
\pp \label{functionF2}
\eea
{}From eq.~\eqn{dzdx}, neglecting the terms proportional to the
derivative of neutrino distributions, one gets the asymptotic value
$z_{eq}^D=(11/4)^{1/3}$ which represents the ratio between the photon
and neutrino temperatures after the complete annihilation of $e^+e^-$
pairs. In the presence of neutrino chemical potentials, the behaviour
of $z(x)$ is in general different. However, the final value of $z(x)$
is always lower than $(11/4)^{1/3}$, showing that also the neutrino
plasma is slightly heated by the $e^+e^-$ annihilations
(mathematically, this can be seen from the fact that the
non-equilibrium contributions involving neutrino distributions in the
r.h.s~of \eqn{dzdx} are of negative definite sign).

In the set of Boltzmann equations \eqn{boltz}, $I_{\nu_\alpha}$ represents
the collisional integral for the single neutrino species $\nu_\alpha$, and
is a functional of all neutrino and $e^{\pm}$ distributions. At the time of
neutrino decoupling, the plasma density was low enough that, in the
expression of $I_{\nu_\alpha}$, one can safely consider only two--body weak
reactions $ 1 + 2 \rightarrow 3 + 4$ with $\nu_\alpha \equiv 1$,
\bea
I_{\nu_\alpha} \left[ f_{\nu_e},f_{\bar{\nu}_e},f_{\nu_x},f_{\bar{\nu}_x}
\right] &=& \frac{1}{2\,E_1} \sum_{\mbox{reactions}} \int \frac{d^3
p_2}{2\,E_2\,(2\,\pi)^3} \: \frac{d^3 p_3}{2\,E_3\,(2\,\pi)^3} \: \frac{d^3
p_4}{2\,E_4\,(2\,\pi)^3} \nonumber \\
&{\times}& (2\,\pi)^4 \, \delta^{(4)} \left( p_1 + p_2 -p_3 - p_4 \right)
\, F \left[ f_1,f_2,f_3,f_4 \right] \left|M_{12\rightarrow34}\right|^2 \vv
\eea
where $F \equiv f_3\,f_4\,\left( 1 - f_1 \right)\,\left( 1 - f_2
\right) - f_1\,f_2\,\left( 1-f_3 \right)\,\left( 1-f_3 \right)$ is the
statistical factor, and $M_{12\rightarrow34}$ is the process
amplitude. In Ref.  \cite{DHS} the complete list of relevant processes
and corresponding squared amplitudes are reported, and it is shown
that using the $\delta$--function some of the integrals can be
analytically performed, reducing $I_{\nu_\alpha}$ to a two-dimensional
integral.

In order to solve eqs.~\eqn{boltz} and \eqn{dzdx}, instead of using a
discretization in momentum space (discrete values for $y$, see
Refs.~\cite{MADSEN,DHS}), we employ a method based on orthonormal
polynomials\footnote{Note that an expansion of the non-equilibrium
distortions in momenta was also discussed in
ref.~\cite{D97}}. According to this general technique, we rewrite the
four unknown neutrino distribution functions as
\be
f_{\nu_\alpha}(x,y) = \frac{1}{e^{y - \xi_\alpha}+1} \left( 1 + \delta
f_{\nu_\alpha}(x,y) \right) \vv
\label{fnu1}
\ee
where $\xi_\alpha$ is the neutrino degeneracy parameter. The initial
conditions on $f_{\nu_\alpha}(x,y)$ are fixed by observing that it is
possible to find a starting value for the evolution parameter
$x=x_{in}$, such that for temperatures larger than the corresponding
$T_{in}$, the neutrino distributions can be safely assumed to be the
equilibrium ones.  This is envisaged by the Fermi factor in the r.h.s
of \eqn{fnu1}, provided a vanishing $\delta f_{\nu_\alpha}$ for
$x=x_{in}$. In this case the initial conditions depend on the two
independent input parameters $\xi_e$ and $\xi_x$ only, being the
antineutrino equilibrium distributions characterized by the corresponding
opposite chemical potentials.  We take $x_{in}= m_e/(10~\mbox{MeV})$
as the starting value for the evolution parameter (as in \cite{DHS}).
Solving the equilibrium part of eq.~(\ref{dzdx}), the initial $z$ is found 
to be $z_{in} = z(x_{in})= 1.00006$.

According to \eqn{fnu1}, the function $\delta f_{\nu_\alpha}$ parameterizes
the departure from equilibrium and can be expanded in terms of a set of
polynomials, $P_i^\alpha(y)$, as
\be
\delta f_{\nu_\alpha}(x,y) = \sum_{i=0}^\infty\, a_{i}^\alpha(x)\,
P_i^\alpha(y) \vv
\label{expan}
\ee
where $P_i^\alpha(y)$ are constructed, with the standard Gram-Schmidt
orthonormalization procedure, and the requirement that they are
orthonormal with respect to the Fermi function weight,
\be
\int_0^\infty \frac{dy}{e^{y - \xi_\alpha}+1} P_i^\alpha(y)
P_j^\alpha(y) = \delta_{ij} \pp
\label{ortho}
\ee
Note that each set of polynomials depends on the neutrino degeneracy
parameter $\xi_\alpha$ through (\ref{ortho}). By substituting
\eqn{expan} in eqs.~\eqn{boltz}, we can rewrite them as 
\be 
\frac{d}{d x}\, a_{i}^\alpha(x) = \frac{1}{x H}\, \int_0^\infty
dy_1\, P_i^\alpha(y_1)\, I_{\nu_\alpha}\left[
f_{\nu_e},f_{\bar{\nu}_e},f_{\nu_x},f_{\bar{\nu}_x} \right] \vv
\label{eqc}
\ee
with $\nu_\alpha=\nu_e, \bar{\nu}_e, \nu_x, \bar{\nu}_x$ and $i=0,1,...$.
{}From the above considerations, the initial conditions for \eqn{eqc}
are $a_{i}^\alpha(x_{in})=0$.
\end{section}

\begin{section}{Numerical details}

In order to solve eqs.~\eqn{dzdx} and \eqn{eqc}, we have to truncate
the infinite series in eq.~\eqn{expan} at a term $i=m$.  The choice of
$m$ is driven by the requested accuracy, which we take to be of the
order $1\%$ in the neutrino distortions. One can make an estimate of
the error in approximating eq.~\eqn{expan} by comparing the results
for two subsequent values $m$ and $m+1$. We have verified that the
requested accuracy can be obtained retaining the coefficients until
$m=3$,
\be
\delta f_{\nu_\alpha}(x,y) \simeq \sum_{i=0}^3\, a_{i}^\alpha(x)\,
P_i^\alpha(y) \pp
\label{texpan1}
\ee
If the expression of each polynomial is considered,
\be
P_i^\alpha(y) = \sum_{j=0}^3~ b_{ij}^\alpha~ y^j \vv
\label{pol}
\ee
one can rewrite Eq. \eqn{texpan1} as
\be
\delta f_{\nu_\alpha}(x,y) \simeq \sum_{i=0}^3\, c^\alpha_{i}(x)\,
y^i \vv
\label{texpan2}
\ee
where the coefficients $c^\alpha_{i}$ are given by
\be
c^\alpha_{i}(x) = \sum_{j=0}^3~ b_{ij}^\alpha~
a_{j}^\alpha(x) \pp
\label{coef}
\ee
The evolution equations for the $n=4 (m+1)+1$ unknown functions
$a_{i}^\alpha$ and $z(x)$ are solved with an integrator for {\it
stiff} equations. This is implemented in the Fortran code by calling a
NAG routine for stiff equations, which uses Backward Differentiation
Formulas with Newton's method and an adaptive step-size. In order to
speed up the evaluation of the r.h.s. of Eq.~\eqn{dzdx}, we made a
fit of the functions $F_1$ and $F_2$ with a precision better than
$1\%$. On the other side, as far as the $4 (m+1)$ equations \eqn{eqc}
are concerned, we needed to compute the tridimensional integral on the
r.h.s.~of eq.~\eqn{eqc}. At this aim we used a routine implementing
the Korobov-Conroy number theoretical method, repeating the
integration twice for calculating the errors, and made a check on the
results by comparing them with the ones obtained with other
integrators.

Actually, one of the equations, that is the energy conservation law
\eqn{dzdx}, is not stiff. Thus in order to check our results for some
values of the neutrino chemical potentials, we used two versions of
the code: in the first one we solved straightforwardly eqs.~\eqn{dzdx}
and \eqn{eqc} as previously described, whereas in the second one we
first separately considered the equilibrium component in $z(x)$, which
satisfies
\be
\frac{d}{dx}z_{eq} = \Frac{\frac{x}{z} F_1 (x/z)}
{\frac{x^2}{z^2} F_1 (x/z) + F_2 (x/z) + \frac{2\pi^4}{15}} \pp
\label{dzeqdx}
\ee
In Figure \ref{f:eqz}, the quantity $z_{eq}$ is plotted versus $x$. It is
independent on the neutrino chemical potentials and asymptotically gives
the value $z^D_{eq}=(11/4)^{1/3}$. Then, the remaining non-equilibrium
correction $\Delta z(x)\equiv z(x)-z_{eq}(x)$ satisfies the differential
equation
\be
\frac{d}{dx} \Delta z= \Frac{- \frac{1}{4z^3} \int_0^\infty~
dy~ y^3~ \left (\frac{d f_{\nu_e}}{dx}+ \frac{d f_{\bar{\nu}_e}}{dx}+ 2\,
\frac{d f_{\nu_x}}{dx}+ 2\, \frac{d f_{\bar{\nu}_x}}{dx} \right )}
{\frac{x^2}{z^2} F_1 (x/z) + F_2 (x/z) + \frac{2\pi^4}{15}} \vv
\label{ddzdx}
\ee 
which has a vanishing r.h.s.~once the neutrinos completely decouple
and their distribution functions get frozen ($df/dx=0$). The two
methods have produced very well coinciding results, and this is a
check of the good performance of the NAG integrators.

\end{section}

\begin{section}{Results}

In order to compare our results with the analysis of Ref.~\cite{DHS},
we have first solved Eqs.~\eqn{dzdx} and \eqn{eqc} for vanishing
neutrino chemical potentials ($\xi_e=\xi_x=0$). In Figures
\ref{f:coefe0} and \ref{f:coefx0} the distortion coefficients
$c^e_{i}$ and $c^x_{i}$, defined in eq.~\eqn{coef}, as a function of
the {\it time} variable $x$ are reported. Note that the antineutrino
distributions coincide with the neutrino ones since we have vanishing
chemical potentials. {}From the plots one can see that the coefficient
evolution ends at $x \sim 10$, which means $T_\gamma = m_e z(x)/x
\sim 0.07$ MeV. After this value the coefficients $c^\alpha_{i}$ reach
their asymptotical values and the neutrino distributions can be well
defined as {\it thermodynamically decoupled} from the electromagnetic
plasma. Note that in both cases, the value of the coefficient
$c^\alpha_3$ is always much smaller than the others, justifying the
truncation of the expansion in \eqn{texpan2} (in reasonable agreement
with the estimate of \cite{DF}).

In Figure \ref{f:evoldf0} the total distortions $\delta f_{\nu_e}$ and
$\delta f_{\nu_x}$ as a function of $x$ are reported, for three values
of neutrino momentum $y$. Note that such distortions increase with the
neutrino energy and, as expected, they are more relevant for electron
neutrinos, due to the fact that only these interact with $e^{\pm}$
through charge currents.  This is clear from Figure \ref{f:df0}, where
we can see that the final distortion $\delta f_{\nu_\alpha}$ is larger
for $\nu_e$ than for $\nu_x$ and is an increasing function of neutrino
energy. The contribution of the distortion to the differential energy
density is given in Figure \ref{f:drho0}, where one can see that the
maximum is for $y\simeq 5$. All the results obtained here for the
non-degenerate case are in perfect agreement with the ones reported in
\cite{DHS}.

In Figure \ref{f:zzeq} we plot the evolution of $\Delta z / z_{eq}$ in
$\%$ with $x$ as obtained from the solution of eq.~(\ref{ddzdx}) for
different choices of neutrino degeneracy parameters. In all cases, it
is a negative decreasing function approaching a constant value when
neutrinos decouple. Thus, from this plot, one can see that the
neutrinos are completely decoupled \footnote{Note that the {\it bump}
at $x \sim3-10$ is only due to the different values of $z$ and
$z_{eq}$ at the late stages of their evolution, and would not appear
if $z(x)$ and $z_{eq}(x)$ were plotted separately.} (i.e.~the
distortions $\delta f$ are frozen) at $x\simeq 3$. One also notices
that $\Delta z$ (the energy transfer from $e^+e^-$ to neutrinos)
slightly diminishes when the neutrino degeneracies increase. This is a
consequence of the fact that the rate of the $e^+e^-$ annihilation
channel into neutrinos is reduced in the presence of neutrino
degeneracy.

In Figures \ref{f:evoldf051A} and \ref{f:evoldf051B} we instead report
the evolution of the total distortions for all neutrino and
antineutrino species as a function of $x$ for given values of the
neutrino degeneracy parameters ($\xi_e=0.5$, $\xi_x=1$) and for three
values of neutrino momentum $y$. Comparing these results with the ones
for the non-degenerate case, we observe that the distortions are
enhanced for antineutrinos and depressed for neutrinos. The different
behaviour between $\nu$ and $\bar{\nu}$ is also evident if one
considers an effective Fermi distribution with a degeneracy parameter
$\xi_{eff}$ as a function of the momentum. It is easy to check that,
at first order in the neutrino distortion, $\xi_{eff}$ is given by
\bea
\xi_{eff}^\nu (x,y) \simeq \xi^\nu + (1 + \exp(\xi^\nu-y)) \delta f_{\nu} (x,y)
\nonumber \\
\xi_{eff}^{\bar{\nu}} (x,y) \simeq -\xi^\nu +
(1 + \exp(-\xi^\nu-y)) \delta f_{\bar{\nu}} (x,y)
\label{xieff}
\eea
In Figures \ref{f:xiefe} and \ref{f:xiefx} we plot the absolute values
of $\xi_{eff}$ for a fixed momentum $y=5$. The total final distortions
and their effect on the differential energy density are plotted in
Figures \ref{f:df051} and \ref{f:drho051}. One can see from the last
plot that the maxima of the neutrino distortions are displaced from
$y\simeq 5$ by an approximate factor $\pm \xi_{\nu_\alpha}$.

We have solved the differential  equations \eqn{dzdx} and \eqn{eqc} for $0
\leq \xi_{\nu_e} \leq 0.5$ and $0 \leq \xi_{\nu_x} \leq 1$ obtaining, for all these
values, the final distribution functions $f_{\nu_{\alpha}}^D(y)$. The
general expression for this quantity as a function of the neutrino degeneracy
parameters, in the above ranges, has been fitted with the following form:
\be
f_{\nu_{\alpha}}^D(y) = \frac{1}{e^{y - \xi_\alpha}+1} \left( 1 +
\sum_{i=0}^3 \sum_{j,k=0}^4 A_i^{\nu_\alpha}(j,k)\, \xi_e^j \, \xi_x^k\,
y^i \exp\left\{ B_i^{\nu_\alpha}\, \xi_e + C_i^{\nu_\alpha}\,
\xi_x\right\}\right) \pp
\label{par}
\ee
The coefficients of these fits are reported in Tables
\ref{t:fitcoefe}--\ref{t:fitcoefbc}. We have checked that in order to
obtain the extension of \eqn{par} for negative values of
$\xi_{\alpha}$ is a good approximation to exchange the neutrino
distribution functions with those of antineutrinos. An example is
given in Figures \ref{f:evoldfe} and \ref{f:evoldfax} for the cases
$\xi_x = \pm 0.6$, both calculated solving the differential equations
\eqn{dzdx} and \eqn{eqc}.

\end{section}

\begin{section}{Phenomenological implications of the neutrino distortions}

There are two cosmological scenarios where the non-equilibrium effects
on the neutrino spectra could in principle manifest: Primordial
Nucleosynthesis and the spectrum of CMBR anisotropies.

We have calculated the change in the production of primordial elements
when the evolution of the neutrino distortions is included. The
non-equilibrium neutrino distributions produce three different
effects.  The first one is to increase the total energy density of the
Universe during BBN and this produces a larger $^4$He abundance. The
inclusion of $\delta f_{\nu_\alpha}$ also modifies the weak $n
\leftrightarrow p$ processes, essentially increasing the rates, which
has as an overall effect the destruction of some neutrons (and a
smaller $^4$He abundance), since they are more abundant than protons.
Finally the evolution law of the photon temperature is also changed to
take into account the small energy transfer from $e^+e^-$ to
neutrinos. Since the final value of the photon temperature is slightly
smaller than in the standard case, this last effect leads also to a
less effective production of $^4$He. In order to facilitate the
comparison of our results with those of previous works, we have
incorporated the distortions $\delta f_{\nu_\alpha} (x,y)$ to the
classic BBN code of Kawano \cite{Kawano}. We have found that the three
effects of the neutrino distortions almost cancel, and the final
change in the mass fraction ($Y$) of $^4$He is only at the $10^{-4}$
level, in good agreement with previous analyses \cite{MADSEN,DHS,BST},
which is undetectable given the current observational
uncertainties. In Figure \ref{f:axis} we show $\Delta Y$ as a function
of the neutrino degeneracies (the changes in the other elements are
even smaller).

The non-equilibrium effects that we have calculated will slightly
modify the relic value of the neutrino energy density. We show in
Figure \ref{f:dnu} the increase of the neutrino energy density,
parameterized in terms of an excess in the effective number of
neutrinos $N_\nu$ as in eq.~\eqn{neff},
$$
\rho_\nu - \rho_\nu^{eq} = \Delta N_\nu \, \frac{7}{4} \,
\frac{\pi^2}{30} \, T_\nu^4
$$
This small increase in $\rho_\nu$ can have a significant effect on the
CMBR. The epoch of matter-radiation equality is very sensitive to the
radiation energy density of the Universe, and the slight heating of
neutrinos changes the predicted spectrum of CMBR anisotropies. Even if
the effect is too small to be noticed with the present data, it has to
be taken into account when determining the other cosmological
parameters \cite{LDHT}. It has been found that the small correction to
the present neutrino energy density is marginally detectable if the
anisotropy and polarization of the CMBR is measured with the expected
precision for the satellite mission Planck \cite{LDHT}.

\end{section}

\begin{section}{Conclusions}

We have calculated the exact kinetic evolution of neutrinos in the
early Universe until complete decoupling, extending previous analyses
to the case in which a large neutrino asymmetry exists. We numerically
solved the Boltzmann equations for the neutrino distribution
functions, finding the momentum-dependent corrections to the
equilibrium spectrum of neutrinos. At the same time, we found the
final value of the ratio between photon and neutrino temperatures,
which is slightly smaller than in the standard case due to small
energy transfer from the electromagnetic plasma to neutrinos due to
$e^+e^- \rightarrow \nu \bar{\nu}$ processes.

Our results are in nice agreement with those of
refs.~\cite{DHS,replyDHS} for the non-degenerate case, obtained
with a different method of solving numerically the kinetic
equations. Instead of a discretization of the distribution function in
momenta, we employed an expansion in momenta of the non-equilibrium part,
truncated at order $y^3$ for an accuracy of $1 \%$. Let us present our
results for this case: $z = 1.39905$ for the final photon temperature,
$\delta \rho_{\nu_e}/\rho_{\nu_e}=0.953~\%$ and $\delta
\rho_{\nu_\mu}/\rho_{\nu_\mu}=0.399~\%$ for the corrections to the
energy density of electron and muon/tau neutrinos, respectively.

When the neutrino degeneracies are non-zero, we found that the
distortions in the neutrino and antineutrino momentum spectra are
different, and larger for antineutrinos (neutrinos) for positive
(negative) asymmetries. We have obtained the final distribution
functions in the range of neutrino degeneracies $0 \leq \xi_{\nu_e}
\leq 0.5$ and $0 \leq \xi_{\nu_x} \leq 1$, and presented the
corresponding fits in Tables \ref{t:fitcoefe}--\ref{t:fitcoefbc}.

We have finally discussed two cosmological scenarios where the
existence of distortions to the spectra of cosmic neutrinos could have
phenomenological implications.  For what concerns Primordial
Nucleosynthesis, we have found that the various effects of the
distortions almost cancel, leading to increases in the final mass
fraction of $^4$He of the order $10^{-4}$ with a small dependence on
the neutrino degeneracies, well below the present observational
uncertainties, in agreement with previous studies.  Finally the
distortions make a small positive contribution $\delta \rho_\nu$ to
the relic value of the neutrino energy density, which slightly
decreases in the presence of neutrino degeneracies. This contribution
increases the radiation content of the Universe and can have an effect
on the CMBR that must be taken into account when determining the other
cosmological parameters \cite{LDHT}.

\end{section}

\section*{Acknowledgements}

The authors thank Alexander Dolgov for fruitful discussions and
encouragement. S.P.~was supported by the TMR network grant
ERBFMRXCT960090.  S.P.~and M.P.~were supported by I.N.F.N.

\begin{table}
\begin{center}
\begin{tabular}{|cc|r|r|r|r|}
\hline\hline
$j$ & $k$  & $i=0$ & $i=1$ & $i=2$ & $i=3$
\\
\hline\hline
   $0$ & $0$ & $-0.02553$ & $-0.02742$ & $0.0613$  & $-0.00148$
\\ $0$ & $1$ & $-0.03074$ & $0.09055$  & $0.1187$  & $-0.00385$
\\ $0$ & $2$ & $-0.02399$ & $-0.12646$  & $0.1148$  & $-0.00271$
\\ $0$ & $3$ & $0.00993$  & $0.08959$  & $0.0496$  & $-0.00505$
\\ $0$ & $4$ & $-0.00835$ & $-0.02629$ & $0.0500$    & $-0.00240$
\\ $1$ & $0$ & $0.15567$  & $-0.26435$ & $0.3008$  & $-0.00832$
\\ $1$ & $1$ & $0.17914$  & $0.94075$   & $0.0113$  & $0.02173$
\\ $1$ & $2$ & $0.18841$  & $1.6935$    & $3.0263$  & $-0.21256$
\\ $1$ & $3$ & $-0.11087$ & $1.5932$    & $-2.8577$ & $0.24631$
\\ $1$ & $4$ & $0.02723$  & $-0.58620$ & $1.4382$  & $-0.13254$
\\ $2$ & $0$ & $-0.40884$ & $-1.4464$  & $1.1112$  & $-0.03258$
\\ $2$ & $1$ & $-0.44811$ & $2.4330$   & $11.350$  & $-0.74164$
\\ $2$ & $2$ & $-0.58457$ & $1.9514$    & $-41.698$ & $3.1148$
\\ $2$ & $3$ & $0.14559$  & $-6.3494$  & $64.550$  & $-4.9244$
\\ $2$ & $4$ & $0.29394$  & $3.3575$   & $-27.901$ & $2.2084$
\\ $3$ & $0$ & $0.53673$  & $0.88730$  & $-0.4563$ & $0.01508$
\\ $3$ & $1$ & $0.60942$  & $1.6758$   & $-38.547$ & $2.9535$
\\ $3$ & $2$ & $0.58380$   & $-15.225$   & $190.83$   & $-15.036$
\\ $3$ & $3$ & $0.86444$  & $23.582$    & $-295.87$  & $24.047$
\\ $3$ & $4$ & $-1.5265$ & $-10.938$   & $141.24$   & $-11.816$
\\ $4$ & $0$ & $-0.28813$ & $-8.3610$    & $5.7948$  & $-0.20675$
\\ $4$ & $1$ & $-0.38332$ & $19.873$    & $53.972$  & $-3.9248$
\\ $4$ & $2$ & $-0.38332$ & $-14.305$    & $-208.88$  & $17.168$
\\ $4$ & $3$ & $-1.5869$ & $-1.7584$  & $351.64$   & $-29.033$
\\ $4$ & $4$ & $1.7923$  & $4.1868$   & $-163.85$  & $13.631$
\\
\hline
\hline
\end{tabular}
\end{center}
\bigskip
\caption{Values of the coefficients $10^{2} {\cdot} A_i^{\nu_e}(j,k)$.}
\label{t:fitcoefe}
\end{table}
%\newpage
\begin{table}
\begin{center}
\begin{tabular}{|cc|r|r|r|r|}
\hline\hline
$j$ & $k$ & $i=0$ & $i=1$ & $i=2$ & $i=3$
\\
\hline\hline
   $0$ & $0$ & $-0.02583$ & $-0.0270$  & $0.0612$   & $-0.001473$
\\ $0$ & $1$ & $-0.07963$ & $-0.0925$ & $-0.13583$ & $-0.003438$
\\ $0$ & $2$ & $-0.19747$ & $-0.1836$ & $0.13921$  & $-0.006921$
\\ $0$ & $3$ & $0.10388$  & $-0.0596$ & $-0.07767$ & $0.002688$
\\ $0$ & $4$ & $-0.38097$ & $-0.2018$ & $0.01913$  & $-0.007743$
\\ $1$ & $0$ & $0.1593$   & $0.0815$  & $-0.00312$ & $0.005457$
\\ $1$ & $1$ & $0.55275$  & $2.8083$  & $-0.07579$ & $0.006372$
\\ $1$ & $2$ & $0.72928$  & $-17.257$ & $0.36785$  & $0.077571$
\\ $1$ & $3$ & $0.02571$  & $35.200$     & $-0.52872$ & $-0.12317$
\\ $1$ & $4$ & $2.0129$  & $-19.508$ & $0.23881$  & $0.098611$
\\ $2$ & $0$ & $-0.42554$ & $0.4269$  & $-0.03213$ & $-0.008874$
\\ $2$ & $1$ & $-1.9241$ & $-34.408$  & $0.93314$  & $0.017523$
\\ $2$ & $2$ & $2.0039$   & $253.21$   & $-4.0506$  & $-0.39445$
\\ $2$ & $3$ & $-7.1596$ & $-508.30$  & $5.7437$  & $0.81968$
\\ $2$ & $4$ & $-1.1437$ & $310.97$   & $-2.5930$ & $-0.55169$
\\ $3$ & $0$ & $0.58393$  & $0.6891$  & $0.07955$  & $0.007278$
\\ $3$ & $1$ & $3.7370$  & $130.57$   & $-0.02633$ & $-0.053902$
\\ $3$ & $2$ & $-12.534$  & $-923.56$  & $11.969$   & $0.77523$
\\ $3$ & $3$ & $28.494$   & $1889.8$    & $-17.361$  & $-1.8111$
\\ $3$ & $4$ & $-9.6482$ & $-1133.0$  & $7.9446$  & $1.2011$
\\ $4$ & $0$ & $-0.33342$ & $1.7667$  & $-0.06659$ & $-0.002354$
\\ $4$ & $1$ & $-2.9576$ & $-137.06$  & $2.3217$  & $0.034407$
\\ $4$ & $2$ & $14.374$   & $1057.1$   & $-10.975$  & $-0.50053$
\\ $4$ & $3$ & $-30.386$  & $-2145.0$  & $16.238$   & $1.2960$
\\ $4$ & $4$ & $13.947$   & $1323.8$   & $-7.5198$ & $-0.87897$
\\
\hline\hline
\end{tabular}
\end{center}
\bigskip
\caption{Values of the coefficients $10^{2} {\cdot}A_i^{\bar{\nu}_e}(j,k)$.}
\label{t:fitcoefae}
\end{table}
%\newpage
\begin{table}
\begin{center}
\begin{tabular}{|cc|r|r|r|r|}
\hline\hline
$j$ & $k$ & $i=0$ & $i=1$ & $i=2$ & $i=3$
\\
\hline\hline
   $0$ & $0$ & $-0.02048$ & $-0.0226$ & $0.03145$  & $-0.00113$
\\ $0$ & $1$ & $-0.00816$ & $-0.121$  & $0.07636$  & $-0.00360$
\\ $0$ & $2$ & $0.02952$  & $-0.1641$ & $0.08906$  & $-0.00585$
\\ $0$ & $3$ & $-0.01427$  & $-0.0915$ & $0.0388$   & $-0.00121$
\\ $0$ & $4$ & $0.00164$  & $-0.1115$ & $0.06871$  & $-0.00862$
\\ $1$ & $0$ & $0.10202$  & $-0.1019$ & $-0.05819$ & $0.00520$
\\ $1$ & $1$ & $0.01809$  & $-0.1574$ & $-0.13567$ & $0.01819$
\\ $1$ & $2$ & $-0.03122$ & $-1.7927$ & $-0.26456$ & $0.01774$
\\ $1$ & $3$ & $-0.10827$ & $0.1559$  & $0.17303$  & $0.02818$
\\ $1$ & $4$ & $0.07637$  & $-0.2702$ & $-0.26288$ & $0.02266$
\\ $2$ & $0$ & $-0.23508$ & $-0.1848$ & $0.04808$  & $-0.01103$
\\ $2$ & $1$ & $0.10452$  & $-7.3509$ & $0.20488$  & $-0.05013$
\\ $2$ & $2$ & $-0.72088$ & $18.548$  & $0.58328$  & $0.02313$
\\ $2$ & $3$ & $1.4737$   & $-21.042$ & $-1.3891$ & $-0.21736$
\\ $2$ & $4$ & $-0.74609$ & $31.385$  & $0.94944$  & $0.07007$
\\ $3$ & $0$ & $0.28818$  & $-0.4307$ & $-0.00637$ & $0.01223$
\\ $3$ & $1$ & $-0.42478$ & $24.659$  & $-0.78479$ & $0.09185$
\\ $3$ & $2$ & $2.5423$   & $-95.411$ & $0.94741$  & $-0.20551$
\\ $3$ & $3$ & $-4.3721$  & $109.17$   & $0.56259$  & $0.67354$
\\ $3$ & $4$ & $2.1039$   & $-36.302$ & $-0.64878$ & $-0.38348$
\\ $4$ & $0$ & $-0.15066$ & $-0.4019$ & $-0.01961$ & $-0.00563$
\\ $4$ & $1$ & $0.40797$  & $-32.973$ & $1.0519$  & $-0.07554$
\\ $4$ & $2$ & $-2.3936$  & $97.440$  & $-2.6746$  & $0.25515$
\\ $4$ & $3$ & $3.9369$   & $-127.23$  & $2.1778$  & $-0.67542$
\\ $4$ & $4$ & $-1.8635$  & $36.555$  & $-0.80063$ & $0.42456$
\\
\hline\hline
\end{tabular}
\end{center}
\bigskip
\caption{Values of the coefficients $10^{2} {\cdot}A_i^{\nu_x}(j,k)$.}
\label{t:fitcoefx}
\end{table}
%\newpage
\begin{table}
\begin{center}
\begin{tabular}{|cc|r|r|r|r|}
\hline\hline
$j$ & $k$ & $i=0$ & $i=1$ & $i=2$ & $i=3$
\\
\hline\hline
   $0$ & $0$ & $-0.02042$ & $-0.0226$ & $0.0315$  & $-0.00113$
\\ $0$ & $1$ & $0.01559$  & $0.0401$  & $0.071$   & $-0.00421$
\\ $0$ & $2$ & $0.02578$  & $0.1409$  & $0.079$   & $-0.00977$
\\ $0$ & $3$ & $0.00504$  & $0.0229$  & $0.0326$  & $-0.00147$
\\ $0$ & $4$ & $-0.00042$ & $0.1407$& $0.0444$  & $-0.01848$
\\ $1$ & $0$ & $0.15327$   & $-0.1203$ & $0.1222$  & $0.00205$
\\ $1$ & $1$ & $-0.11371$ & $1.325$   & $-0.0573$ & $0.00577$
\\ $1$ & $2$ & $-0.19499$ & $-5.7122$ & $1.9867$  & $0.04055$
\\ $1$ & $3$ & $-0.06009$ & $11.340$  & $-2.3154$ & $-0.06816$
\\ $1$ & $4$ & $0.02246$  & $-5.3401$ & $1.2701$  & $0.08527$
\\ $2$ & $0$ & $-0.49132$ & $-0.8147$ & $0.2621$  & $-0.0016$
\\ $2$ & $1$ & $0.33071$  & $-14.744$ & $6.0747$  & $-0.02339$
\\ $2$ & $2$ & $0.67541$  & $100.37$   & $-28.004$ & $0.01113$
\\ $2$ & $3$ & $0.24575$  & $-164.39$   & $45.705$   & $0.11598$
\\ $2$ & $4$ & $-0.13844$ & $95.594$  & $-21.9132$ & $-0.17379$
\\ $3$ & $0$ & $0.76303$  & $1.1339$  & $0.1916$  & $-0.00013$
\\ $3$ & $1$ & $-0.43814$ & $62.116$  & $-22.855$ & $0.15731$
\\ $3$ & $2$ & $-1.1622$  & $-390.91$  & $124.88$   & $-0.75599$
\\ $3$ & $3$ & $-0.45742$ & $662.84$   & $-202.18$  & $1.0609$
\\ $3$ & $4$ & $0.32338$  & $-373.06$  & $102.07$   & $-0.4143$
\\ $4$ & $0$ & $-0.46614$ & $-5.7445$ & $0.6584$  & $0.00111$
\\ $4$ & $1$ & $0.21885$  & $-6.2130$ & $28.568$  & $-0.22558$
\\ $4$ & $2$ & $0.77295$  & $480.42$   & $-143.87$  & $1.23273$
\\ $4$ & $3$ & $0.34072$  & $-752.91$  & $240.49$   & $-2.0296$
\\ $4$ & $4$ & $-0.26958$ & $458.28$   & $-120.81$  & $1.0202$
\\
\hline\hline
\end{tabular}
\end{center}
\bigskip
\caption{Values of the coefficients $10^{2} {\cdot}A_i^{\bar{\nu}_x}(j,k)$.}
\label{t:fitcoefax}
\end{table}

%\newpage
\begin{table}
\begin{center}
\begin{tabular}{|cc|r|r|}
\hline\hline
${\nu_\alpha}$ & $i$ & $B_i^{\nu_\alpha}$ & $C_i^{\nu_\alpha}$ \\
\hline\hline
$\nu_e$ & $0$ & $7.447$ & $-1.268$ \\
$\nu_e$ & $1$ & $-4.521$ & $2.781$ \\
$\nu_e$ & $2$ & $-5.129$ & $-2.015$ \\
$\nu_e$ & $3$ & $-5.405$ & $-2.514$ \\
\hline
$\overline{\nu}_e$ & $0$ & $5.011$ & $-3.179$ \\
$\overline{\nu}_e$ & $1$ & $-3.079$ & $-3.144$ \\
$\overline{\nu}_e$ & $2$ & $0.0847$ & $2.287$ \\
$\overline{\nu}_e$ & $3$ & $3.353$ & $-2.473$ \\
\hline
$\nu_x$ & $0$ & $5.07$ & $0.7437$ \\
$\nu_x$ & $1$ & $-4.445$ & $-1.975$ \\
$\nu_x$ & $2$ & $1.858$ & $-2.301$ \\
$\nu_x$ & $3$ & $4.689$ & $-2.862$ \\
\hline
$\overline{\nu}_x$ & $0$ & $8.037$ & $-0.3521$ \\
$\overline{\nu}_x$ & $1$ & $-6.341$ & $-1.438$ \\
$\overline{\nu}_x$ & $2$ & $-3.938$ & $-2.365$ \\
$\overline{\nu}_x$ & $3$ & $1.812$ & $-4.078$ \\
\hline
\hline
\end{tabular}
\end{center}
\bigskip
\caption{Values of the coefficients $B_i^{\nu_\alpha}$ and $C_i^{\nu_\alpha}$.}
\label{t:fitcoefbc}
\end{table}
\newpage
\begin{center}
\begin{figure}
\psfig{file=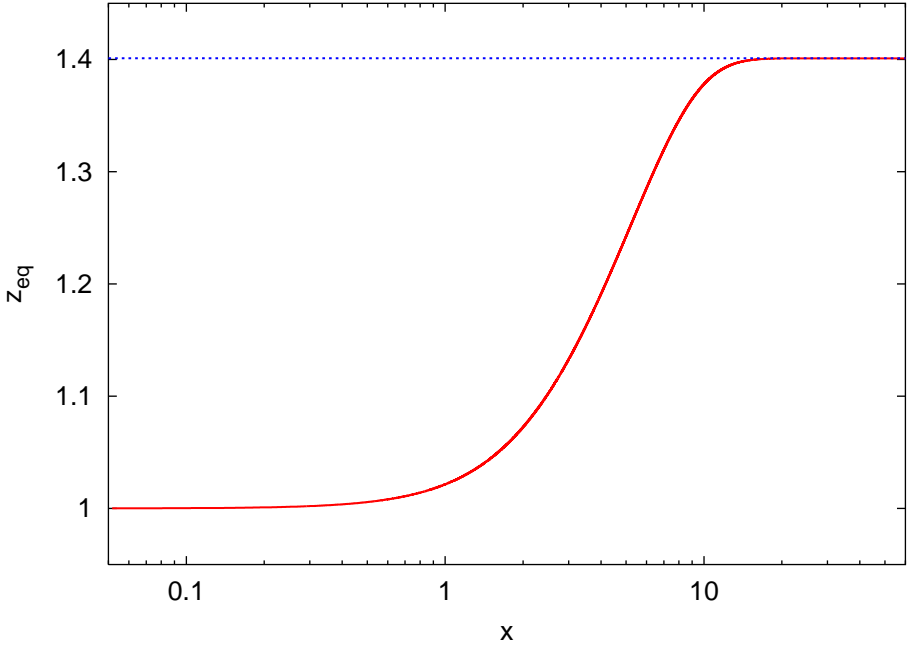,angle=-90,width=.8\textwidth}
\caption{Evolution of the quantity $z_{eq} (x)$, defined in eq.~\eqn{dzeqdx}.
The asymptotical value represents the well-known value
$z^D_{eq}=(11/4)^{1/3}$.}
\label{f:eqz}
\end{figure}
\begin{figure}
\psfig{file=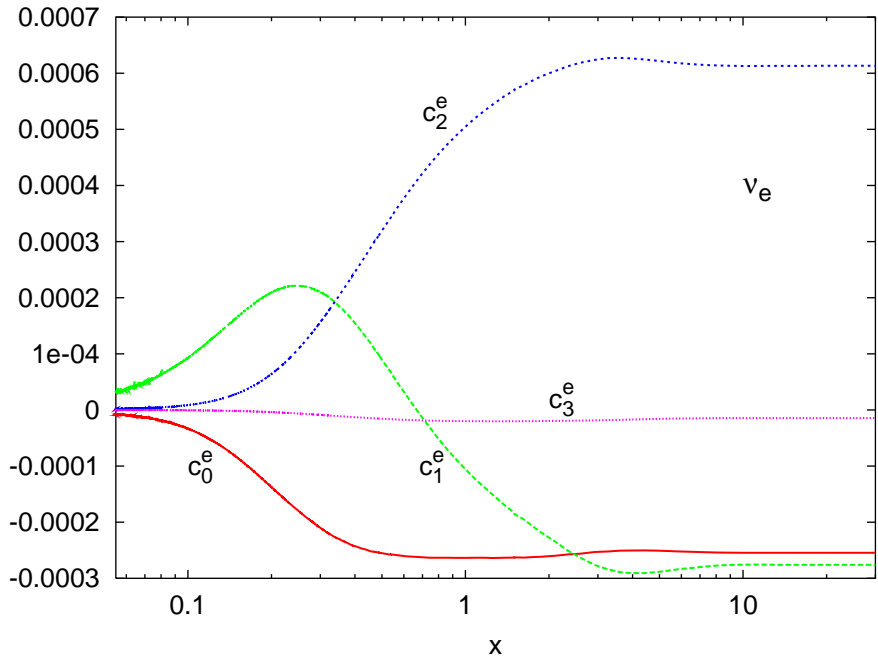,angle=-90,width=.8\textwidth}
\caption{Evolution of the coefficients $c^e_{i}$, as a
function of $x$ (non-degenerate case).}
\label{f:coefe0}
\end{figure}
\begin{figure}
\psfig{file=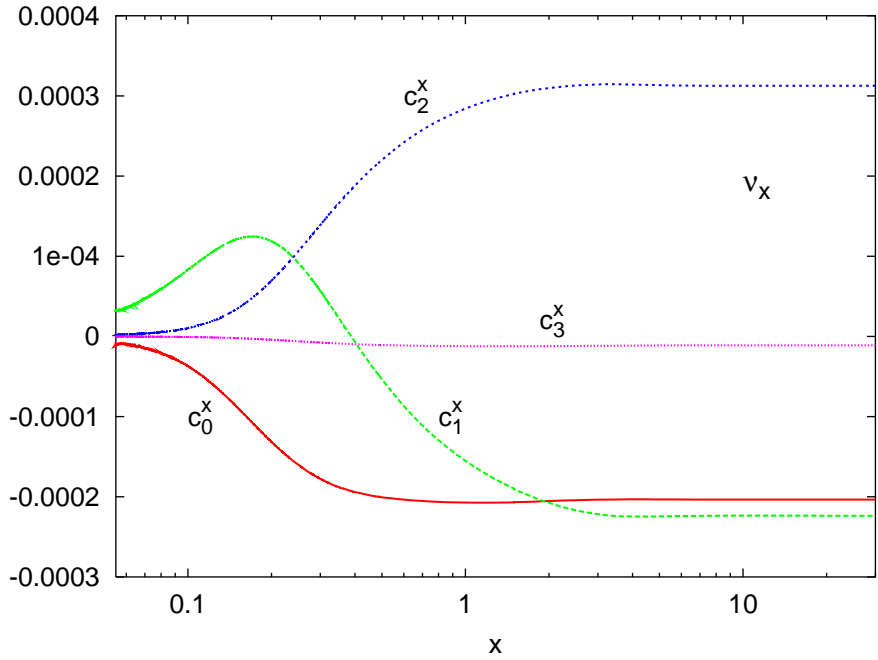,angle=-90,width=.8\textwidth}
\caption{Same as previous figure, for $c^x_{i}$.}
\label{f:coefx0}
\end{figure}
\begin{figure}
\psfig{file=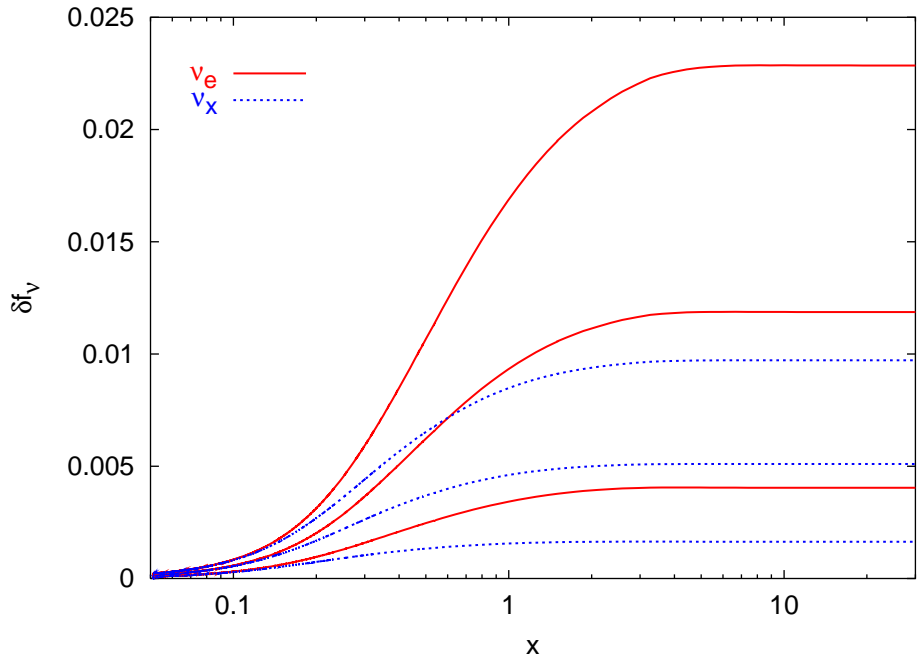,angle=-90,width=.8\textwidth}
\caption{The evolution of the total distortion $\delta f_{\nu_\alpha}$
\eqn{texpan2}, as a function of $x$ (non-degenerate case) for three
values of neutrino momentum $y$. {}From bottom to top: $y=3$, $y=5$
and $y=7$.}
\label{f:evoldf0}
\end{figure}
\begin{figure}
\psfig{file=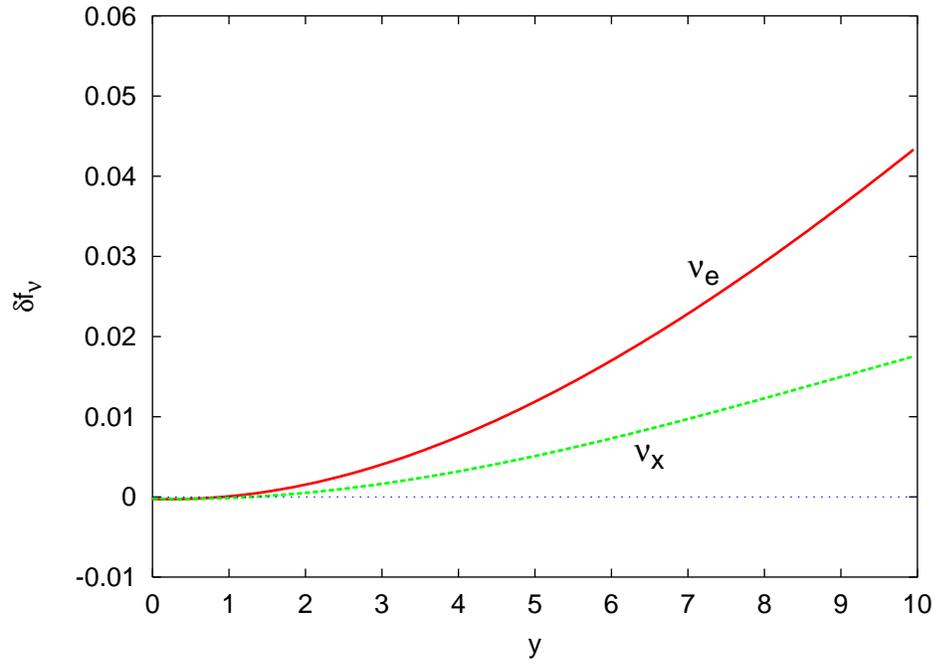,angle=-90,width=.8\textwidth}
\caption{The final distortion as a function of momentum for the
non-degenerate case.}
\label{f:df0}
\end{figure}
\begin{figure}
\psfig{file=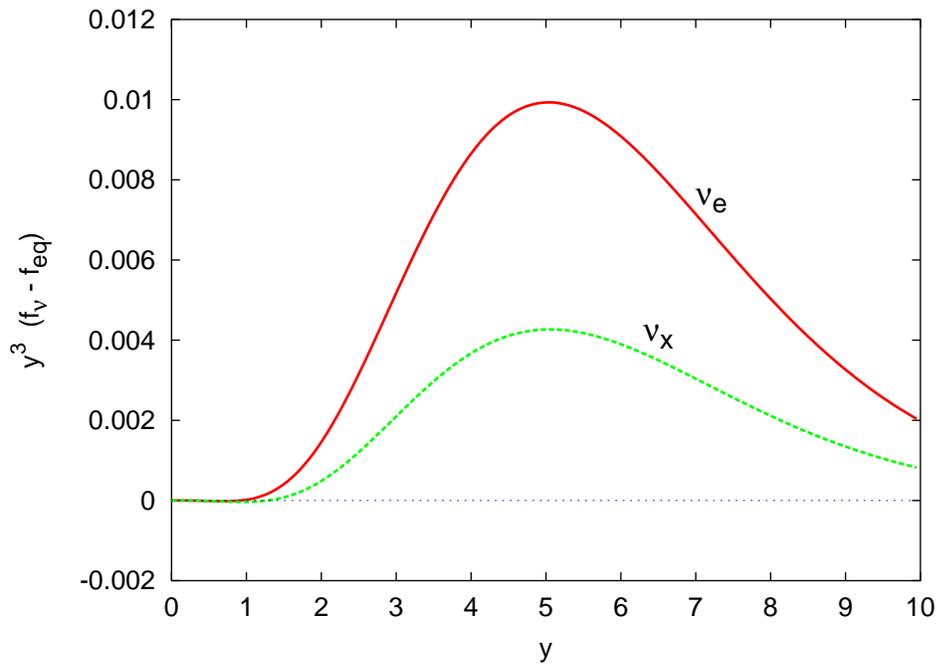,angle=-90,width=.8\textwidth}
\caption{Final distortion of the differential energy
density, as a function of momentum (non-degenerate case).}
\label{f:drho0}
\end{figure}
\begin{figure}
\psfig{file=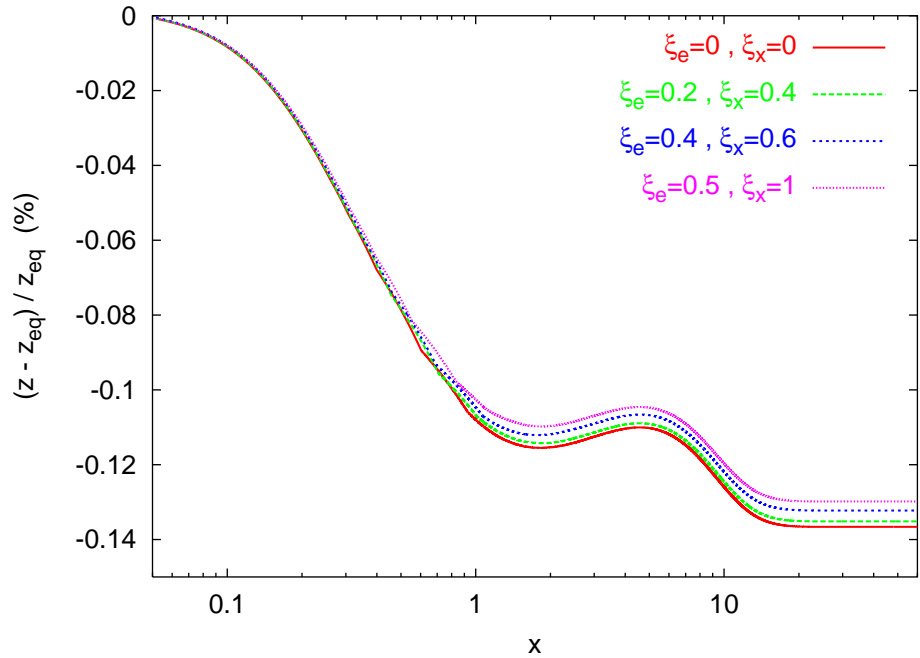,angle=-90,width=.8\textwidth}
\caption{The evolution of $\Delta z/z_{eq}$ from eqs.~\eqn{dzeqdx} and
\eqn{ddzdx}, as a function of $x$ for four choices of the neutrino
degeneracy parameters $\xi_e$, $\xi_x$.}
\label{f:zzeq}
\end{figure}
\begin{figure}
\psfig{file=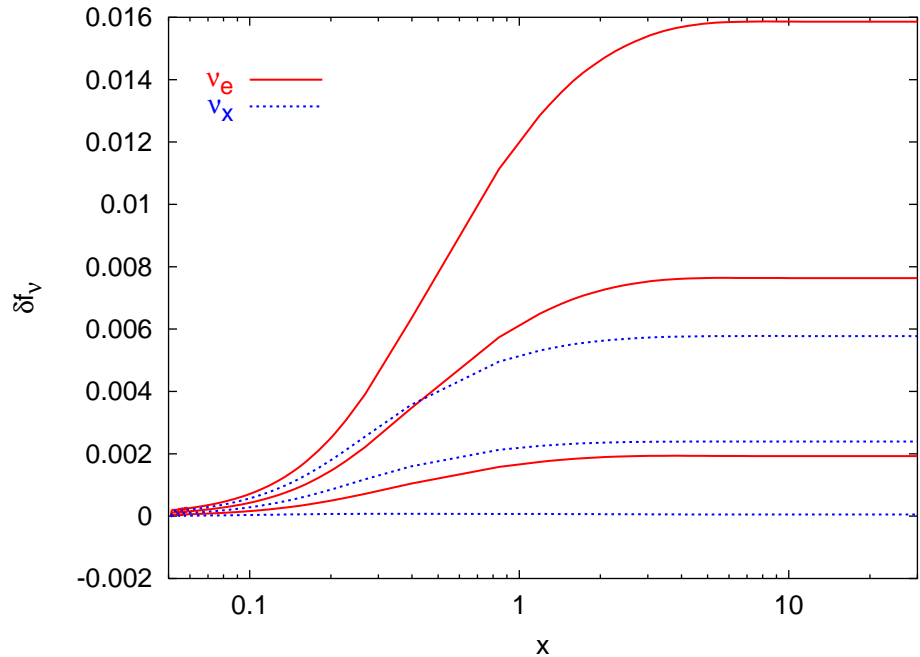,angle=-90,width=.8\textwidth}
\caption{The evolution of the total distortion $\delta f_{\nu_\alpha}$, as a
function of $x$ ($\xi_e=0.5$, $\xi_x=1$) for three values of neutrino
momentum $y$. {}From bottom to top: $y=3$, $y=5$ and $y=7$.}
\label{f:evoldf051A}
\end{figure}
\begin{figure}
\psfig{file=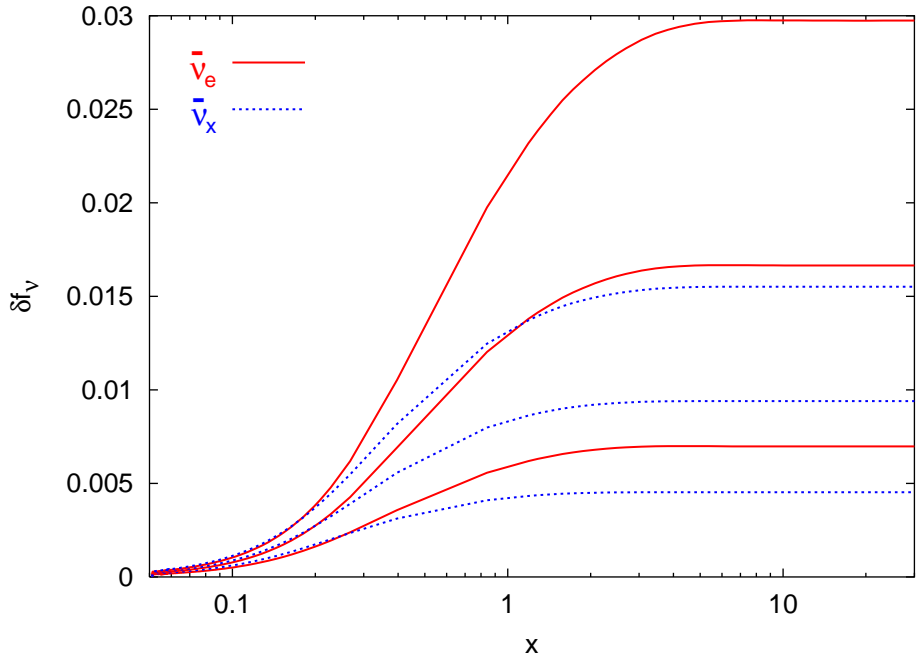,angle=-90,width=.8\textwidth}
\caption{Same as previous figure, for the
antineutrino distortions.}
\label{f:evoldf051B}
\end{figure}
\begin{figure}
\psfig{file=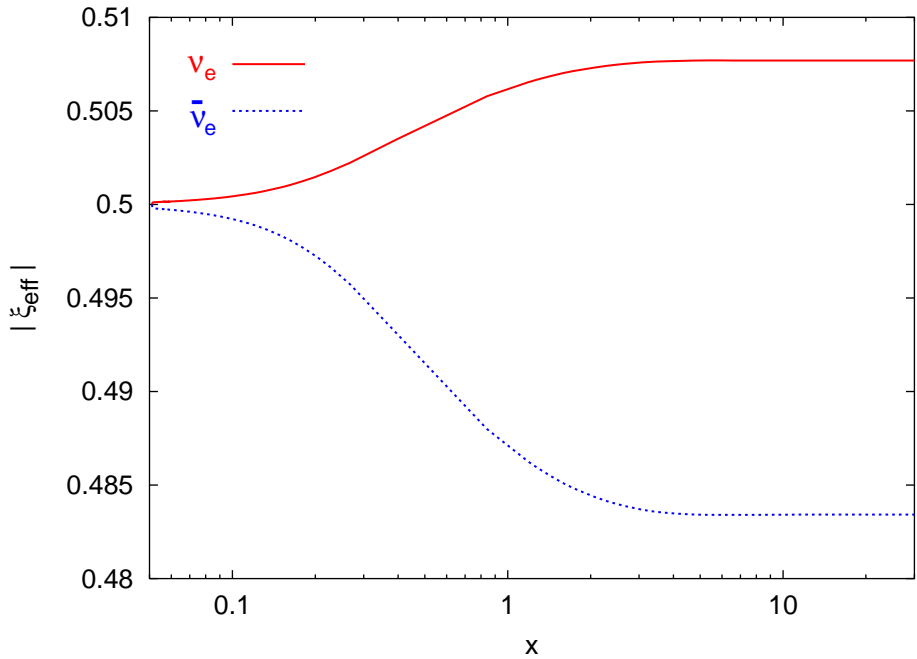,angle=-90,width=.8\textwidth}
\caption{Evolution of the absolute value of
the effective degeneracy parameter $\xi$ (if the distribution function is
written with an equilibrium form) for electron neutrinos and antineutrinos
with momentum $y=5$ ($\xi_e=0.5 , \xi_x=1$).}
\label{f:xiefe}
\end{figure}
\begin{figure}
\psfig{file=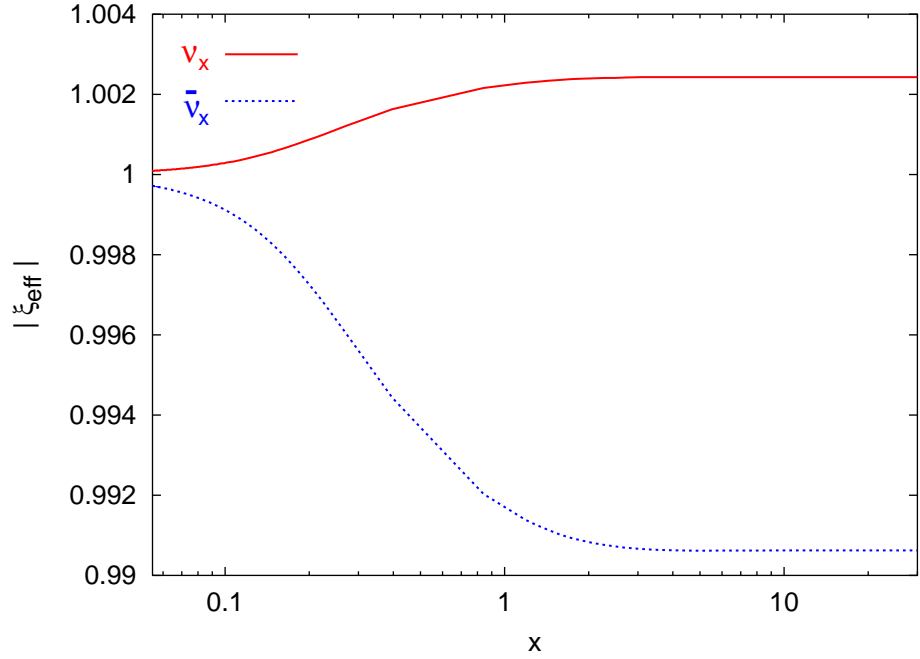,angle=-90,width=.8\textwidth}
\caption{Same as previous figure, for muon or tau neutrinos
and antineutrinos.}
\label{f:xiefx}
\end{figure}
\begin{figure}
\psfig{file=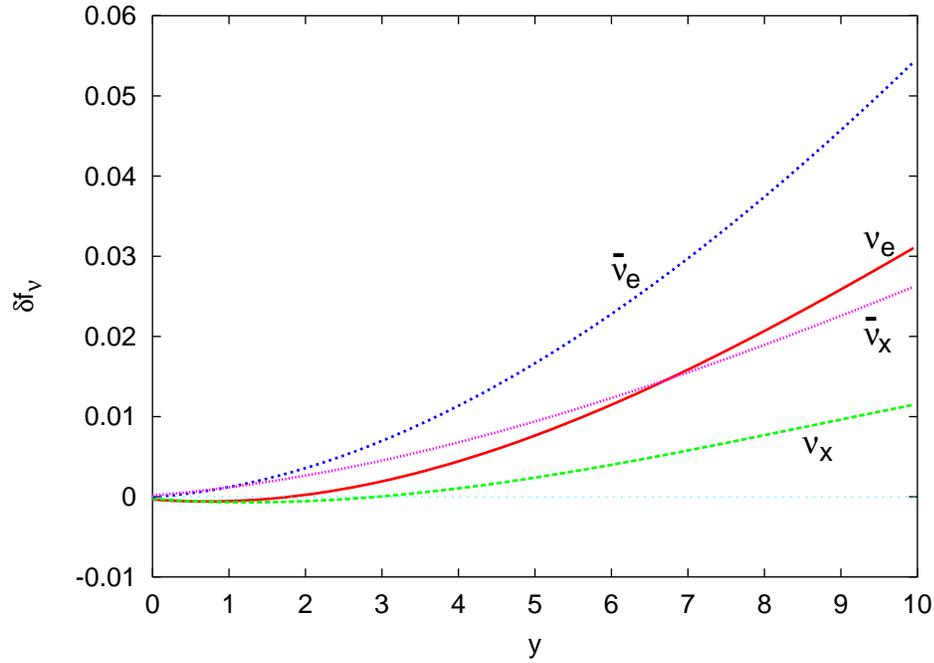,angle=-90,width=.8\textwidth}
\caption{The final distortion as a function of momentum for the case
$\xi_e=0.5$ and $\xi_x=1$.}
\label{f:df051}
\end{figure}
\begin{figure}
\psfig{file=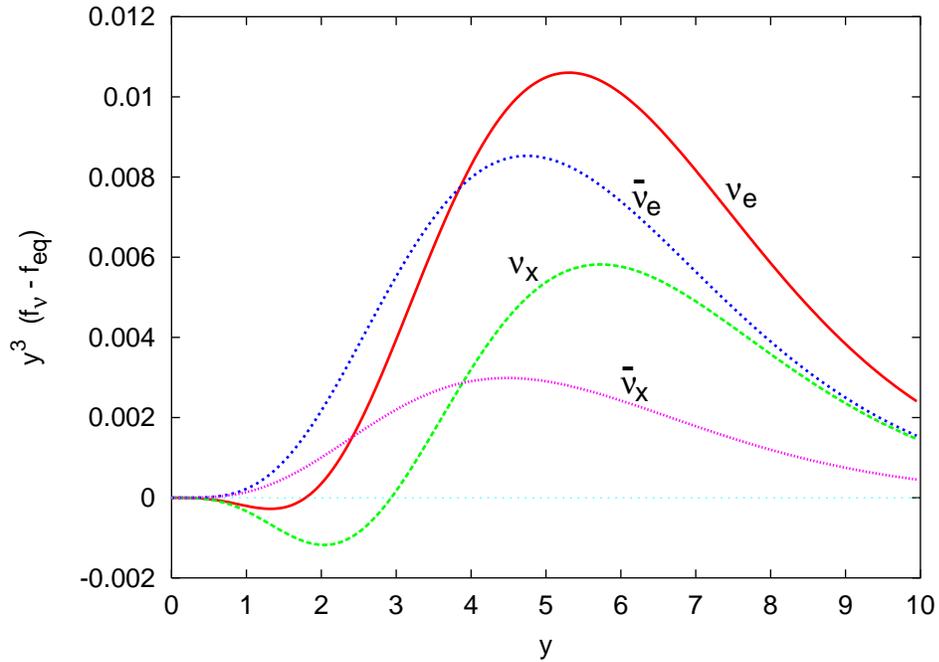,angle=-90,width=.8\textwidth}
\caption{Final distortion of the differential energy
density, as a function of momentum for the case $\xi_e=0.5$ and $\xi_x=1$.}
\label{f:drho051}
\end{figure}
\newpage
\begin{figure}
\psfig{file=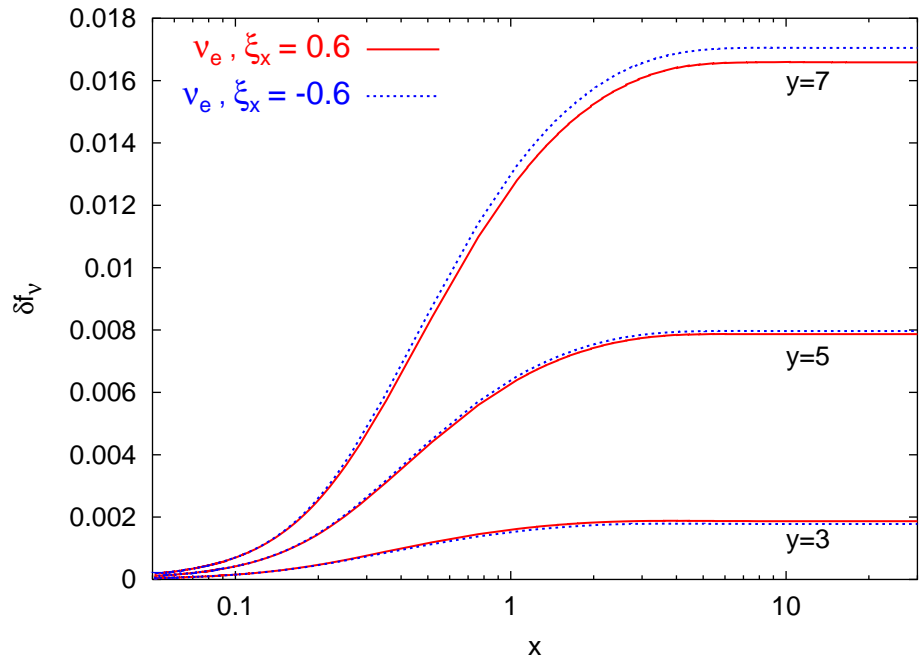,angle=-90,width=.8\textwidth}
\caption{The evolution of $\delta f_{\nu_e}$, as a function of $x$ for
$\xi_e=0.5$, and $\xi_x=0.6$ or $\xi_x=-0.6$, respectively.}
\label{f:evoldfe}
\end{figure}
\begin{figure}
\psfig{file=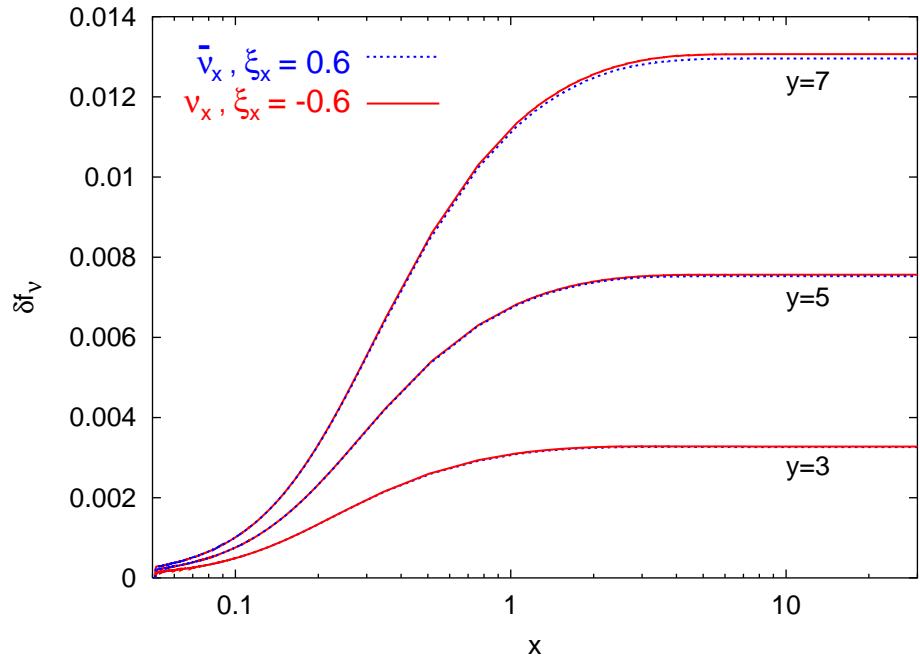,angle=-90,width=.8\textwidth}
\caption{Same as previous figure for $\delta f_{\bar{\nu}_x}$ with
$\xi_x=0.6$ and $\delta f_{\nu_x}$ with $\xi_x=-0.6$.}
\label{f:evoldfax}
\end{figure}
\begin{figure}
\psfig{file=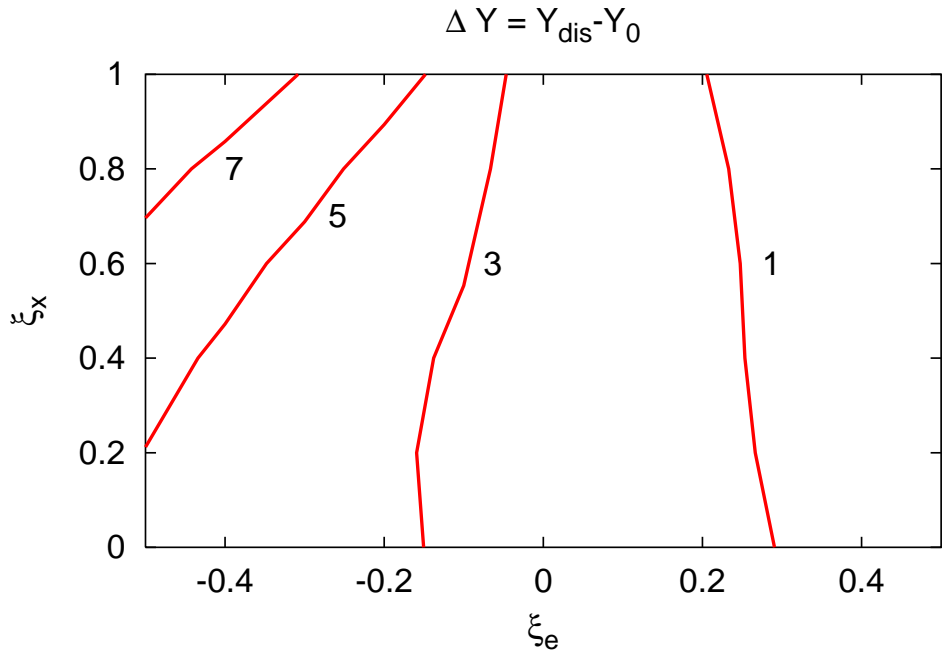,angle=-90,width=.8\textwidth}
\caption{Effect of the neutrino distortion over the
primordial abundance of $^4$He as a function of the neutrino degeneracies
$\xi_e$ and $\xi_x$. The contours indicate equal values of $\Delta Y$ in
units $10^{-4}$.}
\label{f:axis}
\end{figure}
%
%\newpage
\begin{figure}
\centerline{\epsfig{file=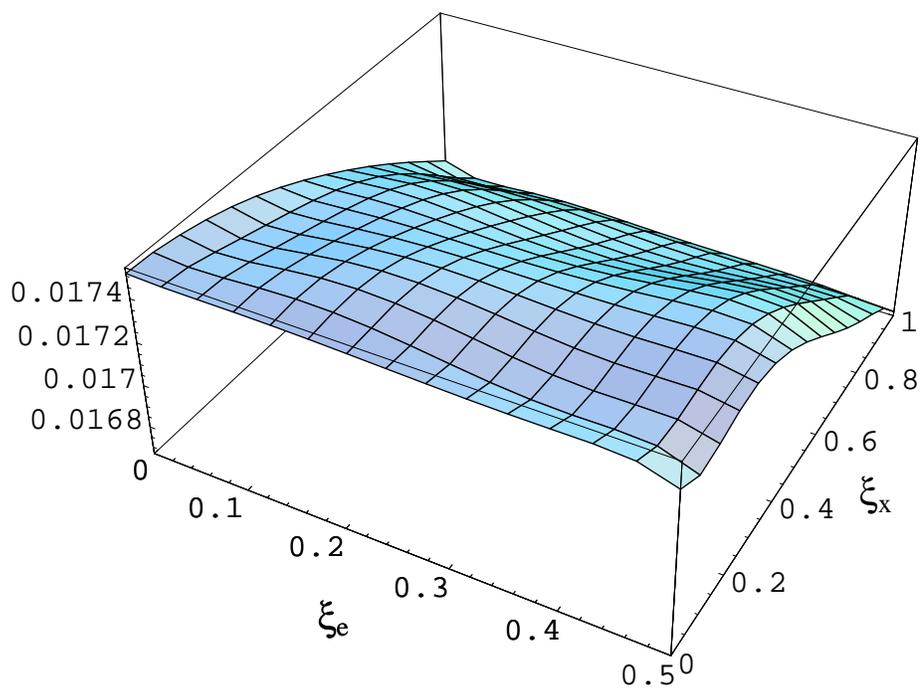,width=.8\textwidth}}
\caption{Effect of the neutrino distortion over the relic neutrino
energy density, parameterized in terms of $\Delta N_\nu$, as a function
of the neutrino degeneracies $\xi_\alpha$.}
\label{f:dnu}
\end{figure}

\end{center}
\end{document}